\documentclass[12pt]{article}
\usepackage[margin=1in]{geometry}
\usepackage{amsmath}
\usepackage{setspace}
\usepackage{graphicx}
\usepackage{epstopdf}
\usepackage[hidelinks]{hyperref}
\usepackage{natbib}
\usepackage{tikz}
\usepackage{authblk}
\usepackage{caption}

\title{G-formula for causal inference via multiple imputation}
\newcommand{\expit}{\mbox{expit}}

\newcommand{\Var}{\mbox{Var}}

\author[1]{Jonathan W. Bartlett}
\author[1]{Camila Olarte Parra}
\author[1]{Emily Granger}
\author[1]{Ruth H. Keogh}
\author[2]{Erik W. van Zwet}
\author[3]{Rhian M. Daniel}
\affil[1]{Department of Medical Statistics, London School of Hygiene \& Tropical Medicine, London, WC1E 7HT, UK}
\affil[2]{Department of Biomedical Data Sciences, Leiden University Medical Center, Leiden, Netherlands}
\affil[3]{Division of Population Medicine, Cardiff University, Cardiff, UK}

\date{}

\begin{document}
\maketitle

\large

\abstract{G-formula is a popular approach for estimating treatment or exposure effects from longitudinal data that are subject to time-varying confounding. G-formula estimation is typically performed by Monte-Carlo simulation, with non-parametric bootstrapping used for inference. We show that G-formula can be implemented by exploiting existing methods for multiple imputation (MI) for synthetic data. This involves using an existing modified version of Rubin's variance estimator. In practice missing data is ubiquitous in longitudinal datasets. We show that such missing data can be readily accommodated as part of the MI procedure when using G-formula, and describe how MI software can be used to implement the approach. We explore its performance using a simulation study and an application from cystic fibrosis.}


\section{Introduction}
The collection of methods referred to as G-methods, developed by James Robins and co-workers, can provide valid inference for the effects of time-varying exposures or treatments in the presence of time-varying confounders---variables that affect treatment over time and the outcome of interest---even when these are affected by previous values of treatment \citep{naimi2017introduction}. One such method is parametric G-formula (sometimes known as G-computation). Parametric G-formula involves postulating models for the time-varying confounders and outcomes. The expected outcome under specified longitudinal treatment regimes of interest can then be estimated and contrasted.

The evaluation of G-formula estimators often involves intractable integrals. To overcome this in practice, G-formula implementations make use of Monte-Carlo integration (simulation) \citep{daniel2011gformula,mcgrath2020gformula}. The Monte-Carlo error in the resulting estimator can be reduced by increasing the number of simulations (simulated individuals). However, the number required to ensure this error is sufficiently small may be quite large - \cite{deStavola2015mediation} found in a data analysis that a Monte-Carlo sample of size 100,000 was required in order for the estimate to be close to a G-formula estimate constructed without the use of simulation. Inference for Monte-Carlo based G-formula estimators is typically performed using bootstrapping. The combination of a large Monte-Carlo sample size and a large number of bootstrap samples can lead to the overall procedure being very computationally intensive. A further common obstacle to using G-formula is that some individuals may have missing values in the time-varying confounders, time-varying treatment variables, or the outcome. In practice, given the already high computational complexity of parametric G-formula, analysts faced with the additional complication of missing data may use simplistic solutions such as last observation carried forward \citep{westreich2012parametric}.

Monte-Carlo simulation-based implementations of G-formula share close connections to the method of multiple imputation (MI) for handling missing data. Previously \citet{westreich2015imputation} 
explained how MI could be used to implement G-formula in a point treatment setting, but stated that Rubin's variance combination rules could not be used here. 
In this paper we show that methods for adapting MI for missing data to generate synthetic datasets can be used to implement a G-formula estimator \citep{Raghunathan2003Varformula}. As such, we show that an accompanying simple variance combination rule can be used to estimate the variance of the resulting estimator, obviating the need for bootstrapping. Furthermore, we show that this MI approach can readily accommodate missing data, thereby providing a convenient single solution to handling the missingness of both actual and counterfactual data.

This paper is organised as follows. In Section \ref{reviewGformula} we review parametric G-formula and how it is typically implemented. In Section \ref{GformViaMI} we describe how G-formula can be implemented by exploiting existing methodology for using MI to generate synthetic datasets. In Section \ref{simulations} we report the results of simulation studies investigating the performance of this approach. Section \ref{illustrativeExample} describes results of an illustrative analysis of data from a cystic fibrosis registry. We conclude in Section \ref{discussion} with a discussion.

\section{Review of G-formula \label{reviewGformula}}
\begin{figure}[hbt!]
    \centering
    \begin{tikzpicture}
    \node (A_0) at (0.00,0.00) {$A_0$};
    \node (A_1) at (3.00,0.00) {$A_1$};
    \node (A_2) at (6.00,0.00) {$A_2$};
    \node (L_0) at (-1.00,2.00) {$L_0$};
    \node (L_1) at (2.00,2.00) {$L_1$};
    \node (L_2) at (5.00,2.00) {$L_2$};
    \node (Y) at (9.00,0.00) {$Y$};
    
    \draw [->] (L_0) edge (L_1);
    \draw [->] (L_0) edge (A_0);
    \draw [->] (L_0) edge (A_1);
    \draw [->] (L_0) edge (A_2);
    \draw [->] (L_0) to[out=90,in=90] (L_2);
    \draw [->] (L_0) to[out=90,in=90] (Y);
    
    \draw [->] (A_0) edge (A_1);
    \draw [->] (A_0) to[out=-90,in=-90] (A_2);
    \draw [->] (A_0) to[out=-90,in=-90] (Y);
    \draw [->] (A_0) edge (L_1);
    \draw [->] (A_0) to[out=90,in=90] (L_2);
    
    \draw [->] (A_1) edge (A_2);
    \draw [->] (A_1) to[out=-90,in=-90] (Y);
    \draw [->] (A_1) edge (L_2);
    
    \draw [->] (A_2) edge (Y);
    
    \draw [->] (L_1) edge (L_2);
    \draw [->] (L_1) edge (A_1);
    \draw [->] (L_1) edge (A_2);
    \draw [->] (L_1) edge (Y);
    
    \draw [->] (L_2) edge (A_2);
    \draw [->] (L_2) edge (Y);
    
    \end{tikzpicture}

    \caption{Directed acyclic graph (DAG) of a study with time-varying treatments $A_0,A_1,A_2$, time-varying confounders $L_0,L_1,L_2$ and final outcome $Y$.}
    \label{fig:causalintroDAG_threetimes}
\end{figure}
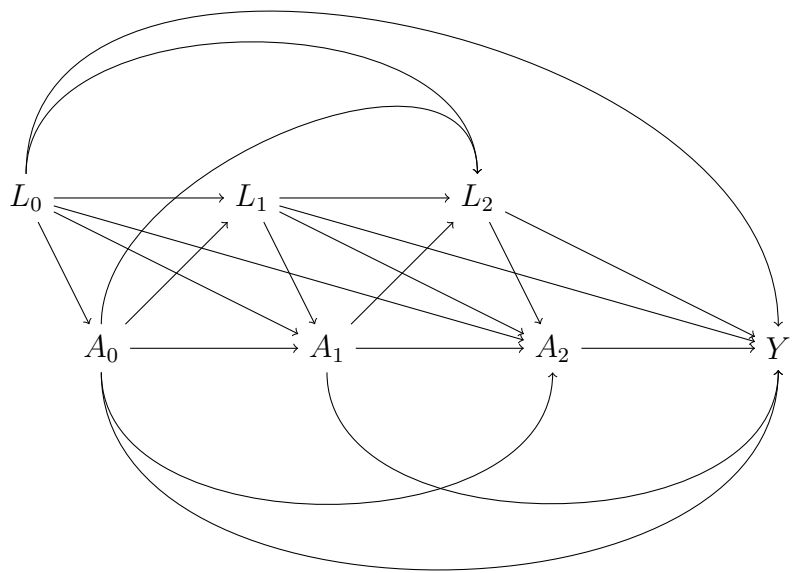

Suppose we have a sample of data from $n_{\text{obs}}$ independent individuals from some well-defined population. For individual $i$ we collect confounder measurements $L_{it}$ at times $t=0,1,\dots,T$. We also collect measurements on the treatment $A_{it}$ at each time, and a final outcome $Y_{i}$. For concreteness, we consider in the following the case of $T=2$, although the developments naturally extend to $T>2$. Figure \ref{fig:causalintroDAG_threetimes} shows a directed acyclic graph (DAG) depicting the assumed causal structure between the variables. We also suppose for now that $L_{it}$ is a real-valued scalar, but natural adaptations/extensions to discrete and higher-dimensional confounders apply. Let $Y^{\bar{a}}$ denote the potential outcome for an individual when their treatment sequence has been set to $\bar{a}$. G-formula relies on certain identification assumptions being satisfied, for the details of which we refer the reader to Chapter 19 of \cite{Hernan2020Ch13}.

The G-formula estimator of $\mu=E(Y^{\bar{a}})=E(Y^{a_0,a_1,a_2})$ for specified values of $a_0$, $a_1$ and $a_2$ is then based on the fact that under the aforementioned identifying assumptions
\begin{align}
     E(Y^{\bar{a}}) &=  \int_{l_0} \int_{l_1} \int_{l_2} E(Y|a_0,a_1,a_2,l_0,l_1,l_2)  f(l_2|a_0,a_1,l_0,l_1)f(l_1|a_0,l_0) f(l_0) dl_2 dl_1 dl_0 
     \label{eq:gform_timevar_exp}.
\end{align}
To implement G-formula we specify and fit models
\begin{align}
    & f(Y|A_0,A_1,A_2,L_0,L_1,L_2; \psi_Y), \nonumber \\
    & f(L_2|A_0,A_1,L_0,L_1; \psi_2), \nonumber \\
    & f(L_1|A_0,L_0;\psi_1), \nonumber \\
    & f(L_0;\psi_0)
    \label{eq:gformModelsNeeded}
\end{align}
and, in principle, evaluate \eqref{eq:gform_timevar_exp} replacing the unknown densities and expectation by their estimated counterparts. However, since \eqref{eq:gform_timevar_exp} cannot generally be evaluated analytically, implementations of G-formula are typically based on Monte-Carlo integration, through simulation of the longitudinal confounders and outcome for each of $n_{\text{syn}}$ individuals, under the treatment combination(s) of interest. That is, given maximum likelihood estimates of the parameters in the conditional models in equation \eqref{eq:gformModelsNeeded}, denoted $\hat{\psi}_0, \hat{\psi}_1, \hat{\psi}_2, \hat{\psi}_Y$, we sequentially simulate for each individual $i=1,\dots,n_{\text{syn}}$ as follows
\begin{align*}
\tilde{L}_{i0} & \sim f(L_0; \hat{\psi}_{0})  \\
\tilde{L}_{i1} & \sim f(L_1| a_0, \tilde{L}_{i0}; \hat{\psi}_{1})   \\
\tilde{L}_{i2} & \sim f(L_2| a_0, a_1, \tilde{L}_{i0}, \tilde{L}_{i1}; \hat{\psi}_{2})   \\
\tilde{Y}_{i}  & \sim f(Y| a_0, a_1, a_2, \tilde{L}_{i0}, \tilde{L}_{i1}, \tilde{L}_{i2}; \hat{\psi}_{Y})  
\end{align*}
The Monte-Carlo G-formula estimator of $E(Y^{a_0,a_1,a_2})$ is then $\frac{1}{n_{\text{syn}}} \sum^{n_{\text{syn}}}_{i=1} \tilde{Y}_{i}$. The number of individuals to simulate for, $n_{\text{syn}}$, could be set equal to $n_{\text{obs}}$, but choosing a larger value reduces Monte-Carlo error in the estimator. For statistical inference, implementations of G-formula in Stata and R rely on the use of non-parametric bootstrapping \citep{daniel2011gformula,mcgrath2020gformula}, which as noted in Section 1, is computationally intensive.

While we have stated that a model $f(L_0;\psi_0)$ is specified and used, in fact this is not needed and is not typically used. Instead, a non-parametric model for $f(L_0)$ is used, and the simulation is performed by sampling a value of $L_0$ from its empirical distribution (that is, sampling $n_{\text{syn}}$ times with replacement from the $n_{\text{obs}}$ observations of $L_0$). Moreover, when as is often the case interest lies in the mean $E(Y^{\bar{a}})$ (as opposed to some other function of the distribution of $Y^{\bar{a}}$), it suffices to specify a model for $E(Y|A_0,A_1,A_2,L_0,L_1,L_2)$, rather than for the full conditional distribution $f(Y|A_0,A_1,A_2,L_0,L_1,L_2)$. Our choice in the preceding to describe a version of G-formula that specifies the latter conditional distribution (rather than mean) model is motivated by the fact this version matches the approach taken in an MI implementation of G-formula, which we describe next.

\section{G-formula via multiple imputation} \label{GformViaMI}
In this section we describe how a Monte-Carlo G-formula estimator can be implemented using MI methods. In Section \ref{gformMIPointEst} we describe how the point estimator is constructed using MI. In Section \ref{gformMIVarEst} we explain why Rubin's standard variance estimator is biased in this instance, and describe an alternative variance estimator, which was derived in the context of using MI to generate synthetic datasets by \citet{Raghunathan2003Varformula}. In Section \ref{gformMISoftware} we describe how standard MI software can be used to implement the approach. Lastly, in Section \ref{gformMIMissingData} we describe how the approach readily extends to accommodate missing actual data (as opposed to missing counterfactual data).

\subsection{Point estimation \label{gformMIPointEst}}
To estimate $E(Y^{\bar{a}})$ by MI, first augment the observed dataset by adding $n_{\text{syn}}$ additional rows. Let $n=n_{\text{obs}}+n_{\text{syn}}$, such that $n$ denotes the number of rows in the augmented dataset (i.e.\ the original plus augmented rows). In the augmented rows, as shown in Table \ref{table:setup}, the baseline and time-varying confounders $(L_0,\dots,L_T)$ and final outcome $Y$  are set to missing, while the treatment variables are set to their values under the regime of interest, i.e. $A_0=a_0, A_1=a_1, \dots, A_T=a_T$. The variable $R$ indicates whether the data row was in the original sample ($R=1$) or not ($R=0$).

Next, Bayesian MI is used to generate $M$ imputations of the missing values in this augmented dataset, using the chosen sequential models (equation \eqref{eq:gformModelsNeeded}). In particular, this means each imputed dataset is generated conditional on a draw from the posterior distribution of the models' parameters. Next, within imputation $m$ ($m=1,\dots,M$), calculate the mean of $Y$ in the augmented rows ($R=0$), yielding $\hat{\mu}_{m} = \frac{\sum^{n}_{i=1} (1-R_i)Y^m_i}{\sum^{n}_{i=1} (1-R_i)}$, where $Y^m_i$ denotes the imputation of $Y_i$ in imputation $m$. The mean outcome under the treatment regime of interest, $\mu=E(Y^{\bar{a}})$, is then estimated as $\hat{\mu} = \frac{1}{M} \sum^{M}_{m=1} \hat{\mu}_{m}$.

The resulting estimator $\hat{\mu}$ differs from the Monte-Carlo G-formula estimator described in Section \ref{reviewGformula} in two respects. First, it generates multiple imputed datasets, analyses each, and combines the estimates, whereas the standard G-formula estimator estimates the mean based on one imputed dataset. Second, whereas the standard G-formula estimator generates the imputed values conditional on efficient (e.g. MLE) estimates of the parameters for the models in equation \eqref{eq:gformModelsNeeded}, as described above the G-formula MI approach generates each imputed dataset conditional on an imputation specific draw from the posterior distribution of the imputation model parameters. These differences are of no consequence for the probability limits of the two estimators --- if we choose $n_{\text{syn}}=k n_{\text{obs}}$ for some fixed $k$, as $n_{\text{obs}} \rightarrow \infty$, the theory for imputation estimators of \cite{Robins/Wang:2000} implies both estimators converge in probability to $\mu=E(Y^{\bar{a}})$ (provided the assumed models are correctly specified).

For the G-formula via MI approach we propose, the impact of generating each imputed dataset conditional on posterior draws of the model parameters, rather than an efficient observed data estimate, is to increase the asymptotic variance of the estimator, but this increase goes to zero as $M \rightarrow \infty$ \citep{Wang1998}. Moreover, this step is essential to facilitate straightforward variance estimation, which we describe next.

\subsection{Variance estimation \label{gformMIVarEst}}
The variance of an MI estimator is typically estimated using Rubin's variance estimator $(1+M^{-1}) \hat{B} + \hat{V}$ where $\hat{B}=\frac{1}{M-1}\sum^{M}_{m=1} (\hat{\mu}_{m}-\hat{\mu})^2$ denotes the between-imputation variance and $\hat{V} = \frac{1}{M} \sum^{M}_{m=1} \widehat{\Var}(\hat{\mu}_{m})$ denotes the average within-imputation variance \citep{Rubin:1987}. While in some settings Rubin's variance estimator is asymptotically unbiased, in some it can be biased upwards or downwards relative to the true repeated sampling variance of the MI point estimator \citep{Robins/Wang:2000}. One such situation is where only a subset of the records used to fit the imputation model is used to fit the analysis model, of which the G-formula via MI estimator is one such example -- the original observed dataset is used to fit the imputation models, while only the augmented dataset rows are used to fit the analysis model (estimating the mean of $Y$ among those with $R=0$). As such we may anticipate that Rubin's variance estimator will be biased for the G-formula via MI point estimator. We demonstrate this empirically in Section \ref{simulations}.

The G-formula via MI estimator is closely related to the use of MI to generate samples from synthetic populations, first proposed by \citet{rubin1993statistical}. Here the objective is to release these synthetic samples rather than the original data in order to protect the confidentiality of survey respondents' data. For synthetic MI,  \citet{Raghunathan2003Varformula} developed $\hat{V}_{\text{syn}} = (1+M^{-1}) \hat{B} - \hat{V}$ as an estimator of $\Var(\hat{\mu})$ from both Bayesian and repeated sampling perspectives.

To build intuition for $\hat{V}_{\text{syn}}$, we now show it is unbiased for $\Var(\hat{\mu})$ in a highly simplified but instructive setting. Suppose we observe data from $n_{\text{obs}}$ individuals on an outcome $Y \sim N(\mu,\sigma^{2})$ and interest lies in inference for $\mu$. Here to estimate the mean $\mu$ we can of course trivially use the sample mean $\bar{Y}=n^{-1}_{\text{obs}}\sum^{n_{\text{obs}}}_{i=1} Y_i$, which has repeated sampling variance $\sigma^2/n_{\text{obs}}$. Suppose however that we use Bayesian MI to generate $M$ new imputed datasets of size $n_{\text{syn}}$. For simplicity, we assume $\sigma^2$ is known. In this case, under the standard non-informative prior for $\mu$, to generate imputation $m$ we first draw  $\tilde{\mu}_{(m)} \sim N(\bar{Y}, \frac{\sigma^{2}}{n_{\text{obs}}})$. For $i=n_{\text{obs}}+1,\dots,n$ we then simulate (impute) $n_{\text{syn}}$ new $Y$ values $Y_{i(m)} = \tilde{\mu}_{(m)} + \epsilon_{i(m)}$, where $\epsilon_{i(m)} \sim N(0,\sigma^{2})$.

Having generated imputed/synthetic datasets for $m=1,\dots,M$, the estimate of $\mu$ based on them is then
\begin{align*}
    \hat{\mu} &= \frac{1}{M} \sum^{M}_{m=1} \hat{\mu}_{m} \\
     &= \frac{1}{M} \sum^{M}_{m=1} \frac{1}{n_{\text{syn}}} \sum^{n}_{i=n_{\text{obs}}+1} \left\{\tilde{\mu}_{(m)} + \epsilon_{i(m)}\right\} \\
     &= \frac{1}{M} \sum^{M}_{m=1} \tilde{\mu}_{(m)} + \frac{1}{n_{\text{syn}}M} \sum^{M}_{m=1} \sum^{n}_{i=n_{\text{obs}}+1}  \epsilon_{i(m)}.
\end{align*}
Letting $\tilde{\mu}=\{\tilde{\mu}_{(1)},\dots,\tilde{\mu}_{(M)}\}$, this has variance
\begin{align*}
    \Var(\hat{\mu}) &= E\left\{\Var(\hat{\mu}|\tilde{\mu}) \right\} + \Var\left\{E(\hat{\mu}|\tilde{\mu}) \right\} \\
    &= E\left\{\Var\left(\left.\frac{1}{n_{\text{syn}}M} \sum^{M}_{m=1} \sum^{n}_{i=n_{\text{obs}}+1}  \epsilon_{i(m)} \right|\tilde{\mu} \right) \right\} + \Var\left\{\frac{1}{M} \sum^{M}_{m=1} \tilde{\mu}_{(m)}\right\} \\
    &= \frac{\sigma^{2}}{n_{\text{syn}}M} + \Var\left\{E\left(\left.\frac{1}{M} \sum^{M}_{m=1} \tilde{\mu}_{(m)}\right|\bar{Y}\right) \right\} + E\left\{\Var\left(\left.\frac{1}{M} \sum^{M}_{m=1} \tilde{\mu}_{(m)}\right|\bar{Y}\right) \right\} \\
    &=  \frac{\sigma^{2}}{n_{\text{syn}}M} + \Var(\bar{Y}) + E\left(\frac{\sigma^2 / n_{\text{obs}}}{M}\right) \\
    &= \frac{\sigma^{2}}{n_{\text{syn}}M} + \left(1+M^{-1}\right)\frac{\sigma^2}{n_{\text{obs}}}.
\end{align*}
With $\sigma^{2}$ known, the within-imputation variance is $\sigma^2/n_{\text{syn}}$ for every imputed dataset, and so $\hat{V}=\sigma^2/n_{\text{syn}}$. Conditional on the observed data $\bar{Y}$, the between-imputation variance estimator $\hat{B}$ is an unbiased estimator of
\begin{align*}
    \Var(\hat{\mu}_{m} | \bar{Y}) &= \Var\left(\left.\tilde{\mu}_{(m)} + \frac{1}{n_{\text{syn}}} \sum^{n}_{i=n_{\text{obs}}+1} \epsilon_{i(m)} \right| \bar{Y}\right) \\
    &= \frac{\sigma^2}{n_{\text{obs}}} + \frac{\sigma^2}{n_{\text{syn}}}.
\end{align*}
Thus, unlike in the missing data setting, the between-imputation variance captures variability both due to uncertainty about $\mu$ in the observed data estimate and the additional variability due to effectively taking new random samples of size $n_{\text{syn}}$ from the population for each imputation \citep{reiter2007multiple}. The expected value of $\hat{V}_{\text{syn}}$ is then
\begin{align*}
    E(\hat{V}_{\text{syn}}) = E\{(1+M^{-1})\hat{B} - \hat{V}\} &= (1+M^{-1}) \left(\frac{\sigma^2}{n_{\text{obs}}} + \frac{\sigma^2}{n_{\text{syn}}}\right) - \frac{\sigma^2}{n_{\text{syn}}} \\
    &=  \frac{\sigma^{2}}{n_{\text{syn}}M} + (1+M^{-1}) \frac{\sigma^2}{n_{\text{obs}}} \\
    &= \Var(\hat{\mu}),
\end{align*}
such that $\hat{V}_{\text{syn}}$ is unbiased for $\Var(\hat{\mu})$. In Web Appendix A we use the results of \citet{Robins/Wang:2000} for the asymptotic behaviour of Rubin's variance estimator to justify $\hat{V}_{\text{syn}}$ more generally.

As noted by \citet{reiter2002satisfying} and \citet{Raghunathan2003Varformula}, the variance estimator $\hat{V}_{\text{syn}}$ can be negative. In the simplified normal mean example, we show in Web Appendix B that the probability of this occurring is approximately given by $    P\left\{ \chi^2_{M-1} < \frac{M}{\frac{n_{\text{syn}}}{n_{\text{obs}}} + 1}  \right\}$. Consideration of this shows, in line with the results of \citet{reiter2002satisfying}, that the probability of a negative variance estimate can be made arbitrarily small by increasing $M$ and/or $n_{\text{syn}}$. \citet{reiter2002satisfying} considers how the latter can be chosen using some initial synthetic imputations to ensure the probability that $\hat{V}_{\text{syn}}$ is negative is sufficiently small. In Section \ref{simulations} we investigate the performance of a procedure where if $\hat{V}_{\text{syn}} \leq 0$, we successively add additional batches of $M$ imputations until  $\hat{V}_{\text{syn}}>0$. To account for the impact of using a finite number of imputations $M$, \cite{Raghu:Rubin:2000} proposed inference based on a t-distribution with degrees of freedom $v_f = (M-1)\left(1-\frac{M\hat{V}}{(M+1)\hat{B}}\right)^2$,
the performance of which we explore in simulations in Section \ref{simulations}.

\subsection{Implementation using imputation software \label{gformMISoftware}}
To implement the proposed approach, as described previously, the observed dataset of size $n_{\text{obs}}$ is augmented by an additional $n_{\text{syn}}$ rows in which all variables are set to missing except the treatment variables, which are set to their values under the regime of interest, i.e. $A_0=a_0, A_1=a_1, \dots, A_k=a_k$. MI software, such as the {\tt mice} package in R, can then be applied to the resulting dataset, with options specified so that the time-varying confounders and outcome are imputed sequentially in time as per the models given in equation \eqref{eq:gformModelsNeeded}. Since the missingness pattern is monotone, no iterative methods such as Markov Chain Monte Carlo are required. Following imputation, the augmented subset is extracted from each imputed dataset, and the mean of $Y$ is evaluated in each, yielding $\hat{\mu}_{m}$ ($m=1,\ldots,M$), along with a corresponding complete data variance estimate. The variance estimator $\hat{V}_{\text{syn}}$ can then be evaluated.

Ordinarily interest focuses on the contrast of potential outcome means under two (or more) different treatment regimes. To estimate the corresponding contrast in potential outcome means, we augment the observed dataset twice. In the second augmentation part, the treatment variables are set according to the second treatment regime of interest. The difference in potential outcome means can be estimated by the difference in simulated outcomes between the two augmented parts. The variance of the resulting estimator can be estimated by the sum of the variance estimator $\hat{V}_{\text{syn}}$ when applied to the two regimes of interest, since the sets of synthetic imputations for the two regimes are independent (conditional on the parameter draws used to impute).

Implementation of the preceding steps using packages such as {\tt mice} in R is relatively straightforward. Nonetheless, to facilitate use of the approach, we provide the R package {\tt gFormulaMI}. This augments the supplied dataset as described above and imputes missing data using the {\tt mice} package. The resulting imputed datasets contain only the augmented portion of the imputations (with $R=0$), which can be used to estimate potential outcome means and contrasts of these. The point estimates and variances from the analysis of these imputations are then passed to a function implementing the variance estimator $\hat{V}_{\text{syn}}$.

As noted earlier, the standard (non-Bayesian) implementation of G-formula avoids specification of a model for $f(L_0)$, and instead simulates from the empirical distribution of $L_0$. This has the advantage of saving the analyst from concerns about misspecification of a model for $L_0$. In the context of MI for generation of synthetic samples, \citet{Raghunathan2003Varformula} proposed using the approximate Bayesian bootstrap approach of \cite{Rubin/Schenker:1987}. In Section \ref{sims:nomissingdata} we investigate in simulations the performance of using this approach for Bayesian non-parametric imputation of $L_0$.

\subsection{Missing data \label{gformMIMissingData}}
Now suppose that there are some data missing. Missing data could occur in either the longitudinal confounders $L_{it}$, the final outcome $Y_i$, or the time-varying treatment variables $A_{it}$. We suppose that the missing at random assumption is deemed plausible for the missing values. In this case, the application of the results of  \citet{Robins/Wang:2000} given in the Web Appendix A still apply without modification. As such, we can impute both the missing data in the original data (where $R=1$) and missing potential outcome data in the augmented rows (where $R=0$), and continue to use the variance estimator $\hat{V}_{\text{syn}}$.

If the missingness pattern in the original data is monotone because of dropout, the missing values in the original data and missing potential outcomes can be imputed simultaneously by imputing sequentially in time, as described in the setting without missing data. More typically, however, the pattern of missingness in the original data will not be monotone. In this case, we propose adapting an approach which works well when the missingness pattern is nearly monotone (Section 6.5.4 of \cite{schafer1997analysis}). We propose that first the missing values in the original data are imputed $M$ times. The augmented rows are then added to each of the $M$ imputed datasets, and the missing potential outcomes in these rows are then imputed once (in each of the $M$ datasets) based on the sequential models (equation \eqref{eq:gformModelsNeeded}).

The models required for G-formula given in equation \eqref{eq:gformModelsNeeded} do not fully specify the joint distribution of all the variables under consideration, since they do not specify models for the treatment variables. The imputation models used to impute the missing data in the original dataset should ideally be compatible with those used to impute the augmented rows. One way to achieve this is to specify a full joint model for all the variables by, in addition to the models in equation \eqref{eq:gformModelsNeeded}, specifying models for the time-varying treatment variables. That is, for $t=0,1,\dots T$, we specify models $f(A_{it}|\bar{A}_{i(t-1)},\bar{L}_{it})$, such as suitable logistic regression models if treatment is binary. While imputation from such a joint model is possible using Bayesian model software such as JAGS, imputation is more commonly performed using methods such as chained equations, as implemented in the popular R package {\tt mice}. As such, in Section \ref{sims:missingdata} we investigate performance when the models used to impute missing data are not strictly compatible with the models specified and used by G-formula (in equation \eqref{eq:gformModelsNeeded}).

In the setting with missing data, our R package {\tt gFormulaMI} takes as input a set of $M$ imputed datasets, for example obtained using the {\tt mice} package. It then augments each imputed dataset with the required additional rows and imputes each dataset once.

\section{Simulations \label{simulations}}
In this section we report the results of simulations performed to examine the empirical performance of the G-formula via MI approach. We first consider, in Section \ref{sims:nomissingdata}, the setting where there is no missing data. Next, in Section \ref{sims:missingdata}, we consider the situation where some data are missing.

\subsection{No missing data} \label{sims:nomissingdata}
We simulated datasets for 500 individuals with a single continuous confounder $L$ measured at times $t=0,1,2$, corresponding binary treatments $A$, and a continuous final outcome $Y$. The specific data generating mechanism is given in Web Appendix C. We report results for estimates of $E(Y^{1,1,1})-E(Y^{0,0,0})$, whose true value under the data generating mechanism is 3. The G-formula via MI approach was implemented using the {\tt mice} package in R, imputing $L_0$, $L_1$, $L_2$ and $Y$ from normal linear models including all the preceding (in time) treatment and confounder variables linearly. Since the missingness pattern is monotone, we specified that {\tt mice} only perform one iteration. We set $n_{\text{syn}}=500$. To investigate how performance varied with $M$, we evaluated the procedure using $M=5,10,25,50,100$. If in a particular simulation $\hat{V}_{\text{syn}}<0$, we added an additional $M$ imputations and re-calculated $\hat{V}_{\text{syn}}$. This was repeated until $\hat{V}_{\text{syn}}>0$.

Table \ref{tab:simsNoMissing} shows results based on 10,000 simulations per scenario. As expected since the imputation models were correctly specified, the G-formula via MI estimator for $E(Y^{1,1,1})-E(Y^{0,0,0})$ was unbiased for all values of $M$. The variance estimator $\hat{V}_{\text{syn}}$ was also essentially unbiased. Confidence intervals calculated based on a t-distribution with degrees of freedom $v_f$ showed overcoverage for $M=5,10,25$, but achieved nominal coverage for $M=50$ and $M=100$. Confidence intervals calculated based on a standard normal showed substantial undercoverage for $M=5$ and $M=10$, but had close to nominal coverage for $M=50$ and $M=100$. Lastly, when using a smaller initial value for $M$, sometimes additional imputations were required to ensure $\hat{V}_{\text{syn}}>0$, as indicated by the mean and maximum $M$ values in Table \ref{tab:simsNoMissing}. However, when an initial $M=50$ (or $M=100$) imputations were used, $\hat{V}_{\text{syn}}$ was always positive.

We additionally ran 10,000 simulations with $M=50$ using the approximate Bayesian bootstrap to impute $L_0$. The variance estimator $\hat{V}_{\text{syn}}$ was again unbiased. The coverage of the confidence interval constructed using a t-distribution with degrees of freedom $v_f$ was 94.9\% while the normal based confidence interval had coverage 93.8\%, matching closely the results of scenario 4 in Table \ref{tab:simsNoMissing}.

\subsection{Missing data} \label{sims:missingdata}
Next we performed simulations where  some data were missing. Data in each of $L_1$, $A_1$, $L_2$, $A_2$ and $Y$ were made missing completely at random, with the probability of each being missing set to $\pi$, with $\pi=\{0.05, 0.1, 0.25, 0.5\}$. As such, the probability of an individual having complete data was $(1-\pi)^5$ and the average number of variables missing per individual was $5\pi$. Thus $\pi=0.5$ is a really quite extreme scenario, with only approximately $3\%$ of individuals having complete data.

To implement G-formula via MI we used an initial call to {\tt mice} to impute the missing values $M=50$ times. The continuous variables $L_1$, $L_2$ and $Y$ were imputed using normal linear models while $A_1$ and $A_2$ were imputed using logistic regression models. Since the missingness pattern was not monotone, as per the standard chained equations algorithm, for imputation of a given variable, all the other variables were included as covariates. The number of iterations was left at its default value of 5, except for $\pi=0.5$. Here, with a very large amount of missingness, we found that 50 iterations were required to achieve convergence. Having imputed the missing data, the additional $n_{\text{syn}}=500$ rows were added to each imputed dataset, and {\tt mice} was applied to each of the $M=50$ datasets, specifying to impute using one iteration sequentially according to time, as used in the scenario without missing data.

Table \ref{tab:simsMissing} shows the results based on 10,000 simulations per value of $\pi$. The G-formula via MI estimator had minimal bias across all four scenarios. As we would expect, the empirical standard error increased with increasing amounts of missing data. The variance estimator $\hat{V}_{\text{syn}}$ was positive in all simulations and for all values of $\pi$ when using an initial value of $M=50$. $\hat{V}_{\text{syn}}$ was unbiased for the empirical SE. Confidence intervals based on a t-distribution with degrees of freedom $v_f$ showed slight overcoverage, while the normal based intervals showed slight undercoverage.

\section{Illustrative example}
\label{illustrativeExample}
In this section we provide an illustrative example of the use of the G-formula via MI approach to investigate the effects of multiple treatments on lung function in people with cystic fibrosis (CF). Many people with CF are prescribed at least one mucoactive treatment to help improve lung function. In the UK, the most commonly prescribed nebulized mucoactive treatment is dornase alfa (DNase), and many patients already using DNase may later add or switch to hypertonic saline. Existing research investigates the effects of taking DNase or hypertonic saline alone, but the effects of using both treatments in combination are less well understood. Here we investigate the following question: for people with CF who are already established on DNase, does adding hypertonic saline have any additional benefit for lung function? In a recent study this question was investigated using marginal structural models estimated using inverse probability of treatment weighting to address time-dependent confounding \citep{granger2023emulated}. 

Our example uses data from the UK Cystic Fibrosis Registry, which collects longitudinal data on almost all people with CF in the UK \citep{taylor2018data}. Longitudinal data are collected annually, when CF patients are seen at an outpatient clinic for a comprehensive review. The review data includes evaluation of clinical status, lung function, chronic medications, hospital admissions and health complications.

Using data from 2007 to 2018, we included individuals with CF, aged 6 years or older, who had been prescribed DNase, but not hypertonic saline, for at least two consecutive years. Organ transplant recipients, and people prescribed certain treatments (mannitol, ivacaftor, lumacaftor/ivacaftor, tezacaftor/ivacaftor) were excluded. Time zero was defined as the date of the most recent annual review at which the inclusion and exclusion criteria were met, but which allowed for the maximum possible follow-up time up to 5 years. The outcome of interest is lung function, and this is measured at the annual review as forced expiratory volume in one second (FEV$_1$\%). We estimate the mean differences in FEV$_1$\% at times 1-5 years had all individuals been prescribed DNase and hypertonic saline, compared to if all individuals were prescribed DNase only. The following variables were considered as confounders: sex, CFTR genotype, ethnicity, date of birth, rate of decline in FEV$_1$\% during the year prior to time 0, past FEV$_1$\%, respiratory infections, IV hospital admissions, BMI, pancreatic insufficiency and use of IV antibiotics. The first five of these were baseline confounders; the latter six were time-varying. Data may be missing if some information is not recorded at the annual review, or if the individual is no longer in the registry due to death or administrative end-of-follow-up. For the purposes of illustration, we assume all such missingness is at random.

Treatment effect estimates were obtained using the standard implementation of G-formula and using G-formula via MI. Standard implementation was done using the R package {\tt gfoRmula}, with $n_{\text{syn}}$ set to $100,000$ as used by \cite{deStavola2015mediation}. Normal-based confidence intervals were constructed based on the non-parametric bootstrap estimated standard error (1,000 bootstrap samples). Implementation of G-formula via MI was done using {\tt mice} and {\tt gFormulaMI}, with $M=200$ and $n_{\text{syn}}=n_{\text{obs}}=4,759$. Since the G-formula MI estimator is the average of the estimates across the $M$ imputations, this yields an effective Monte-Carlo sample size of $200 \times 4,759 \approx 951,800$. Whereas {\tt gFormulaMI} uses multiple imputation to handle missing data (as described in Section \ref{gformMIMissingData}), the {\tt gfoRmula} package handles missing data using a complete case analysis approach. For individuals who had any missing data at a particular time point, their data for that time point was not included in the conditional models used to simulate covariates and outcomes. Another difference between the two packages is the way the conditional models are defined. In the {\tt gfoRmula} package, combined models are fitted across all time points, where time is usually included as a predictor. In the {\tt gFormulaMI} package, separate models are fitted sequentially in time (equation (2)). Consequently, any differences in the results obtained between the two packages could be due to a combination of the different approaches to defining conditional models and the different approaches to handling missing data. To differentiate between these two sources of differences in the results, we also used both packages to analyse a complete dataset. For the complete data, we used the first imputed dataset created by {\tt mice}. For each analysis, we recorded the running time and estimated the Monte Carlo standard error for the treatment effect estimate (formulae for Monte Carlo standard errors are provided in Web Appendix D). 

4,759 individuals were eligible for inclusion in the analysis. Details on how the study sample was selected, baseline characteristics by treatment group, and amount of missing data by year, are provided in Web Appendix D, Web Tables 1 and 2, and Web Figure 1. Table \ref{tab:appliedResults} shows the results from each analysis along with the total running time.  

Overall, and in line with previous results \citep{granger2023emulated}, we found little evidence that adding hypertonic saline has any effect on FEV$_1$\% among individuals who are already established on DNase. Most 95\% confidence intervals contained 0, and for the few cases where 0 lies outside the interval, the estimated effect size was not clinically significant. Results from the two different implementations were broadly speaking similar. However, there were consistent differences in the estimated effect sizes obtained using between the two packages. These differences were similar in the analyses using complete data and those from analyses including missing data, suggesting that the way the conditional models are defined was the main driver of differences between the {\tt gfoRmula} and {\tt gFormulaMI} results, rather than the approach used to handle missing data. This finding is perhaps not surprising given the relatively low amounts of missing data. Compared to the {\tt gfoRmula} package, the {\tt gFormulaImpute} package consistently obtained smaller Monte Carlo standard errors and had considerably shorter running time. 

\section{Discussion}
\label{discussion}
G-formula via MI is an attractive approach for implementing parametric G-formula. With complete data, inference for G-formula estimators is usually based on non-parametric bootstrapping. While the bootstrap provides a consistent variance estimator under mild assumptions, since correct model specification is generally required for consistency of the G-formula point estimator, it makes sense to use a variance estimator (e.g. based on Bayesian MI) which exploits an assumption that these models are correct.

To obtain point estimates and inferences with sufficiently small Monte-Carlo error, existing simulation based implementations of G-formula may require both the Monte-Carlo sample size to be large and also a large number of bootstrap samples to be used. Although our simulations and data analysis are limited in scope, they suggest that reliable inferences can be obtained via MI methods using smaller Monte-Carlo sample sizes ($n_{\text{syn}}$) and relatively few imputations (e.g. 50). Although implementation is relatively straightforward using existing MI packages, we have developed an R package {\tt gFormulaMI} that interfaces with the {\tt mice} package to perform the required data manipulation steps, estimate mean outcomes under each treatment regime of interest, and calculate the synthetic MI variance estimator. Imputation packages such as {\tt mice} are flexible in regard model specification, for example allowing the possibility for the user to include interactions and higher order effects in models.

We note that while standard G-formula is typically implemented using Monte-Carlo simulation as we have described, an alternative version based on iterative conditional expectations can be used \cite{wen2021parametric}. This approach requires models for a series of conditional mean functions, rather than for the full distribution of the time-varying confounders and outcome. This makes it potentially less prone to model misspecification, particularly in the case of several time-varying confounders, where the standard approach requires a choice of factorisation to be made to specify the distribution of the time-varying confounders at each time point. Moreover, it does not require the use of simulation, and closed form variance estimators can be constructed based on estimating equation theory \citep{zivich2023empirical}. 

In practice datasets, whether arising from experimental or observational studies, have missing data to a lesser or greater extent. In this context the G-formula via MI approach has greater appeal, given that it provides a coherent potential solution to handle both the missing actual and missing counterfactual data. While our simulation investigations suggest the G-formula via MI approach can perform well, our conclusions regarding its empirical performance in general are necessarily limited by the fact our simulations have only explored a relatively simple setup - with one continuous confounder and a small number of time points. 

One alternative to imputing (actual) missing data when implementing G-formula is to fit each of the models (in equation \eqref{eq:gformModelsNeeded}) using the subset of records for which the variables involved in each model are fully observed. These complete case model fits yield consistent estimates of the respective conditional model parameters provided the probability of having all the variables involved in the model is independent of the dependent variable conditional on the covariates \citep{white2010bias}. When the pattern of missingness in the longitudinal dataset is complex, consisting of both intermittent missingness and missingness due to dropout, such an assumption can sometimes be deemed more plausible than missing at random, whose meaning becomes complex in such settings \citep{robins1997non}. Thus an alternative possible version of the G-formula via MI approach when some data are missing is to fit each of the required models using their respective complete case fits. Imputations for the potential outcomes in the augmented rows of the dataset could then generated conditional on draws from the posterior distributions of the parameters given these complete case fits.

In this paper we have focused on G-formula where the outcome is a variable $Y$ measured at some final time point, but the approach can also be used for causal inference for outcomes measured at several time points, as in the CF example. Moreover, G-formula can be used when the outcome is the time to some event of interest, for example based on discrete time logistic regression models \citep{westreich2012parametric}. The G-formula via MI approach can also be used in this setting, by defining appropriate time-dependent binary indicators of survival. 

Implementations of G-formula (e.g. the G-formula packages in Stata \citep{daniel2011gformula} and R \citep{mcgrath2020gformula}) often fit models pooled across time points for each variable. This is achieved by formatting the data in so-called long form. Doing so permits borrowing of information across time points in the estimation of regression parameters, but of course relies on the validity of the assumption that the conditional distribution of confounders given earlier variables is homogenous across time points. Although this approach could be implemented via the MI approach we have outlined, we do not believe it is possible using standard imputation software such as {\tt mice} in R. This is because having transformed the data into long form, it is not possible to update values from one row of the data frame from another within the algorithm, which would be required for subsequent times points for the same subject.

While our focus in this paper has been on static treatment regimes, G-formula can be used to estimate the effects of dynamic treatment regimes, where the exposure or treatment at a given time point is assigned dependent on the longitudinal history observed up to that time. The G-formula via MI approach can be extended to this case, by setting the treatment variables to missing in the augmented part of the dataset and then specifying how they should be imputed based on the preceding (in time) variables. This can be achieved for example in the {\tt mice} package through the use of user specified deterministic (or indeed stochastic) imputation methods. Moreover, we believe the basic approach we have described can be extended and applied to more general and complex causal structures specified by a DAG.

\section*{Acknowledgements}
Bartlett, Olarte Parra and Daniel are supported by a UK Medical Research Council Grant (MR/T023953/1). Keogh and Granger are supported by a UKRI Future Leaders fellowship (MR/S017968/1) awarded to Keogh.

The authors thank people with CF and their families for consenting to their data being held in the UK CF Registry, and NHS teams in CF centres and clinics for the input of data into the Registry. We also thank the UK Cystic Fibrosis Trust and the Registry Steering Committee for access to anonymised UK CF Registry data.

\section*{Supporting Information}
Web Appendices 1-4, Tables 1-2, and Figure 1 referenced in Sections 3, 4 and 5 are available in the Supporting information, located later in this document.

R code for the simulation study can be found at \url{https://github.com/jwb133/gFormulaViaMultipleImputation}. The R package {\tt gFormulaMI} is available from CRAN.

\clearpage

\begin{table}
\centering
\caption{G-formula via MI data setup. The original dataset (top part) is augmented with additional rows (bottom part). In the augmented part, confounders $L_0,L_1,L_2$ and outcome $Y$ are set to missing (indicated here by NA), while the treatment variables $A_0,A_1,A_2$ are set to their values under the regime of interest (here $1,1,1$). The variable $R$ indicates whether the row is originally observed data ($R=1$) or not ($R=0$).}
\begin{tabular}{llllllll}
$R$ & $L_0$ & $A_0$ & $L_1$ & $A_1$ & $L_2$ & $A_2$ & $Y$ \\
\hline
1 & -0.3 & 0 & 0.5 & 0 & 2.2 & 1 & 1.3 \\ 
1 & 2.3 & 1 & 4.2 & 1 & 4.6 & 1 & 5.5 \\ 
1 & -0.5 & 1 & 0.4 & 0 & 0.8 & 1 & 1.9 \\ 
1 & -0.1 & 0 & 1.6 & 1 & 4.1 & 0 & 7.0 \\ 
1 & 0.4 & 1 & 1.9 & 1 & 3.5 & 1 & 6.2 \\ 
\hline
0 & NA & 1 & NA & 1 & NA & 1 & NA \\
0 & NA & 1 & NA & 1 & NA & 1 & NA \\
0 & NA & 1 & NA & 1 & NA & 1 & NA \\
0 & NA & 1 & NA & 1 & NA & 1 & NA \\
0 & NA & 1 & NA & 1 & NA & 1 & NA \\
\hline
\end{tabular}
\label{table:setup}
\end{table}

\clearpage

\begin{table}[ht]
\centering
\caption{Simulation results for G-formula via MI without any missing data. Results are shown for different numbers of initial imputations $M$. Emp SE. gives the empirical standard error of the point estimates while Est. SE gives the mean estimated standard error based on $\hat{V}_{\text{syn}}$. Raghu df 95\% CI gives the coverage of t-based confidence intervals based on the degrees of freedom $v_f$ while Z 95\% CI gives coverage for confidence intervals constructed using $N(0,1)$ quantiles. Mean $M$ and Max $M$ give the mean and maximum value of $M$ required across the simulations in order to obtain $\hat{V}_{\text{syn}}>0$.} 
\begin{tabular}{rrrrrrrrr}
  \hline
Scenario & M & Bias & Emp. SE & Est. SE & Raghu df 95\% CI & Z 95\% CI & Mean M & Max M \\ 
  \hline
 1 & 5 & 0.002 & 0.244 & 0.238 & 99.7 & 87.1 & 5.6 & 15 \\ 
 2 & 10 & -0.007 & 0.233 & 0.223 & 98.3 & 89.1 & 10.2 & 30 \\ 
 3 & 25 & 0.002 & 0.223 & 0.219 & 95.5 & 92.7 & 25.0 & 25 \\ 
 4 & 50 & 0.002 & 0.221 & 0.219 & 94.9 & 93.7 & 50.0 & 50 \\ 
 5 & 100 & -0.003 & 0.219 & 0.219 & 95.0 & 94.6 & 100.0 & 100 \\ 
   \hline
\end{tabular}
\label{tab:simsNoMissing}
\end{table}

\clearpage

\begin{table}[ht]
\centering
\caption{Simulation results for G-formula via MI with missing data. $\pi$ is the probability that each of $L_1$, $A_1$, $L_2$, $A_2$ and $Y$ are missing.} 
\begin{tabular}{rrrrrrr}
  \hline
Scenario & $\pi$ & Bias & Emp. SE & Mean est. SE & Raghu df 95\% CI & Z 95\% CI  \\ 
  \hline
 1 & 0.05 & 0.000 & 0.226 & 0.225 & 94.9 & 93.8 \\ 
 2 & 0.10 & -0.005 & 0.232 & 0.232 & 95.2 & 94.2 \\ 
 3 & 0.25 & -0.010 & 0.260 & 0.258 & 95.0 & 94.1 \\ 
 4 & 0.50 & -0.013 & 0.357 & 0.361 & 95.3 & 94.5 \\ 
   \hline
\end{tabular}
\label{tab:simsMissing}
\end{table}

\clearpage

\begin{table}[ht]
\centering
\caption{Results using data from the UK CF Registry. TE: Treatment Effect (estimated effects of adding hypertonic saline on FEV$_1$\% in people already taking DNase); CI: Confidence Interval; MCSE: Monte Carlo Standard Error. Treatment effects for years 1-5 are estimated using two packages: {\tt gfoRmula} and {\tt gFormulaImpute} and using two datasets: original sample (including missing data) and a complete dataset (obtained by imputing the missing values once using {\tt mice}). Times given in hours indicate computational time to run each analysis.}
\begin{tabular}{rrrrrrr}
  \hline
 &  & Year 1 & Year 2 & Year 3 & Year 4 & Year 5  \\ 
  \hline
   & & & & & & \\
  \multicolumn{7}{l}{\textbf{Complete data}} \\
  \multicolumn{7}{l}{{\tt gfoRmula} (122.51 hours)} \\
  & TE & 0.00 & 0.24 & 0.51  & 1.34  & 1.40  \\
 & 95\% CI & (-0.46, 0.47)  & (-0.40, 0.88) & (-0.39, 1.41) & (0.03, 2.64) & (0.15, 2.64)  \\
  & MCSE & 0.101 & 0.105 & 0.110 & 0.107 & 0.098  \\
 & & & & & & \\
 \multicolumn{7}{l}{{\tt gFormulaImpute} (1.36 hours)} \\ 
  & TE & 0.35  & -0.06  & 0.21  & 1.06  & 1.11   \\
  & 95\% CI & (-0.69, 1.39) & (-1.28, 1.17) & (-0.88, 1.30) & (-0.12, 2.25) & (-0.20, 2.41)  \\
   & MCSE & 0.037 & 0.044 & 0.039 & 0.042 & 0.046  \\
  \hline
  \\
  \multicolumn{7}{l}{\textbf{Partially observed dataset}} \\
  \multicolumn{7}{l}{{\tt gfoRmula} (169.61 hours)} \\
    & TE & 0.08  & 0.47 & 0.49  & 1.59  & 1.56 \\
    & 95\% CI & (-0.55, 0.70) & (-0.39, 1.33)  & (-0.79, 1.77) & (-0.35, 3.53) & (-0.47, 3.58)  \\
     & MCSE & 0.104 & 0.110 & 0.112 & 0.109 & 0.096 \\
        & & & & & & \\
    \multicolumn{7}{l}{{\tt gFormulaImpute} (11.14 hours)} \\ 
  & TE  & 0.39 & -0.18  & 0.14  & 1.05 & 1.05   \\
  & 95\% CI & (-0.57, 1.34) & (-1.29, 0.94) & (-1.01, 1.28) & (-0.29, 2.40)  & (-0.33, 2.43)  \\
   & MCSE & 0.034 & 0.039 & 0.041 & 0.048 & 0.049 \\
  \hline
\end{tabular} 
\label{tab:appliedResults}
\end{table}

\clearpage

\large

\begin{center}\textbf{Supporting information for `G-formula for causal inference via multiple imputation' by Jonathan W. Bartlett, Camila Olarte Parra, Emily Granger, Ruth H. Keogh, Erik W. van Zwet, Rhian M. Daniel}
\end{center}

\normalsize

\section*{Web Appendix A}
In this appendix we show that the variance estimator $\hat{V}_{\text{syn}}$ derived by \cite{Raghunathan2003Varformula} is consistent for the G-formula via MI estimator, using the results of \cite{Robins/Wang:2000}. We first describe how the G-formula via MI estimator can be embedded into the setup of \cite{Robins/Wang:2000}. To that end, consider the augmented dataset where in the augmented part $\bar{L}$ and $Y$ are set to missing and the treatment vector $\bar{A}$ is set to different values according to the treatment regimes of interest. Define the variable $R$ that takes value 1 in the original observed data and 0 in the augmented part of the data. The full data vector is thus $F=(\bar{A},\bar{L},Y,R)$ and the observed vector is $O=(\bar{A},\bar{L}R,YR,R)$.

The missing values in $\bar{L}$ and $Y$ in the augmented part of the dataset are then multiply imputed. After imputation, in the G-formula via MI approach we estimate counterfactual means and contrasts of these using only data in the augmented rows, i.e. among those rows with $R=0$. The complete data estimating functions are thus of the form
\begin{equation*}
u(F,\beta) = (1-R)W(\bar{A},\bar{L},Y,\beta),
\end{equation*}
where $\beta$ is the parameter to be estimated. Suppose for example that potential outcomes are imputed for treatment regimes $\bar{a}_1$ and $\bar{a}_2$. Then letting $\beta=(\beta_1,\beta_2)=(E(Y^{\bar{a}_1}),E(Y^{\bar{a}_2}))$, the complete data estimating function could be of the form
\begin{equation}
    u(F,\beta) = (1-R)\left\{Y-\beta_1 I(\bar{A}=\bar{a}_1)-\beta_2 I(\bar{A}=\bar{a}_2)\right\} \begin{pmatrix}
    I(\bar{A}=\bar{a}_1) \\
    I(\bar{A}=\bar{a}_2)
    \end{pmatrix},
    \label{eq:completeDataEstFun}
\end{equation}
which corresponds to estimating the potential outcome mean for each of the two regimes by the sample means in the augmented data among those assigned to each regime.

Under the regularity conditions detailed by \cite{Robins/Wang:2000}, the G-formula via MI estimator of 
$\beta$ which uses $M=\infty$ imputations is asymptotically normal. The variance of the estimator with $M=\infty$ is given by equation A3 of \cite{Robins/Wang:2000} as
\begin{align*}
\Sigma =& \tau^{-1} \Big[ E\left\{U_{obs}(\psi^{*},\beta^{*})^{\otimes 2}\right\} + \kappa \Lambda(\psi^{*}) \kappa^{T} + \kappa E\left\{ D(\psi^{*})U(\psi^{*},\beta^{*})^{T}\right\} \\ 
 &+ E\left\{ D(\psi^{*})U(\psi^{*},\beta^{*})^{T}\right\}^{T} \kappa^{T} \Big] (\tau^{T})^{-1},
\end{align*}
where $A^{\otimes 2}=A A^T$. In the following we define the terms involved in the preceding expression, derive their values in our setting, and show that $\hat{V}_{\text{syn}}$  is a consistent estimator of this asymptotic variance. Parameters with a superscript $*$ denote the true value of the corresponding parameter.

The quantity $\tau$ is defined as
\begin{equation*}
\tau = - E\left\{\frac{\partial U(\psi^{*},\beta)}{\partial \beta^{T}} \right\} \bigg | _{\beta=\beta^{*}},
\end{equation*}
where
\begin{align*}
U(\psi,\beta)=u(F(\psi),\beta).
\end{align*}
Here $F(\psi)$ denotes an imputed full data vector for a given random individual, with the imputation generated conditional on the value $\psi$ of the imputation model parameter. Thus for an individual in the original data ($R=1$), $F(\psi)=(\bar{A},\bar{L},Y,1)$, while for an individual in the augmented part ($R=0$), $F(\psi)=(\bar{A},\bar{L}(\psi),Y(\psi),0)$, where $(\bar{L}(\psi),Y(\psi))$ denotes the random imputed value drawn from $f(\bar{L},Y|\bar{A},\psi)$.

The quantity $U_{obs}(\psi,\mu)$ is defined as $E_{\psi}\{U(\psi,\beta) | O\}$. In our case, we have
\begin{align*}
U_{obs}(\psi,\beta) &= E_{\psi}\{U(\psi,\beta)|O\} \\
&= (1-R)E_{\psi}\{W(\bar{A},\bar{L}(\psi),Y(\psi),\beta) |\bar{A},R=0\} \\
&=0,
\end{align*}
provided $E_{\psi}\{W(\bar{A},\bar{L}(\psi),Y(\psi),\beta)|\bar{A},R=0\}=0$. This holds for example for the complete data estimating function given in equation \eqref{eq:completeDataEstFun}.

Letting $\hat{\psi}$ denote the observed data MLE of the imputation model parameters, $D(\psi)$ denotes the influence function of the estimator, and under standard regularity conditions $n^{1/2}(\hat{\psi}-\psi^{*})$ is asymptotically normal with mean zero and covariance matrix equal to
\begin{align*}
\Lambda(\psi^{*}) = E\left\{D(\psi^{*})^{\otimes 2} \right\} = I^{-1}_{obs} E\left\{S^{\otimes 2}_{obs}(\psi^{*}) \right\} I^{-1}_{obs}
\end{align*}
where $S_{obs}(\psi)$ denotes the observed data score and $I_{obs}$ the observed information matrix. As noted by Robins and Wang following their Theorem 2, if, as we assume, the imputation model is correctly specified, $I_{obs}=E\left\{S^{\otimes 2}_{obs}(\psi^{*}) \right\}$, in which case $\Lambda(\psi^{*})=I^{-1}_{obs}$.

Observations in the augmented dataset, with $R=0$, do not contribute to the estimation of the imputation model, and so for such observations $D(\psi^{*})=0$. Conversely,  observations in the original data with $R=1$ do not contribute to the estimation of $\beta$ in the imputed datasets:
\begin{align*}
U(\psi,\beta)=(1-R)W(\bar{A},\bar{L},Y,\beta)=0 \text{ if } R=1
\end{align*}
Consequently, $D(\psi^{*})U(\psi^{*},\mu^{*})=0$ for all observations, and so 
\begin{align*}
E\left\{ D(\psi^{*})U(\psi^{*},\beta^{*})^{T}\right\}=0.
\end{align*}

Thus the asymptotic variance of the G-formula via MI estimator, with $M=\infty$, is given by
\begin{align}
\Sigma = \tau^{-1} \kappa I^{-1}_{obs} \kappa^{T} (\tau^T)^{-1}
\end{align}
where
\begin{align*}
\kappa = E\{ U(\psi^{*},\beta^{*}) S_{mis}(\psi^{*})^{T} \}, 
\quad S_{mis} = \frac{\partial}{\partial \psi} \log f(F|O;\psi) |_{\psi=\psi^{*}}.
\end{align*}

Equation A1 from Robins and Wang gives that when imputations are generated from a Bayesian model (as in MI as originally conceived by Rubin), the (standardised by $n$) between imputation variance estimator $\bar{B}$ converges in probability as $m,n \rightarrow \infty$ to
\begin{eqnarray*}
\tau^{-1} \left\{\kappa I^{-1}_{obs} \kappa^{T} + E\left[\{U(\psi^{*},\beta^{*})-U_{obs}(\psi^{*},\beta^{*})\}^{\otimes 2}\right]  \right\} (\tau^T)^{-1} \\ 
= \tau^{-1} \left\{\kappa I^{-1}_{obs} \kappa^{T} + E\left[U(\psi^{*},\beta^{*})^{\otimes 2}\right]  \right\}(\tau^T)^{-1},
\end{eqnarray*}
since $U_{obs}(\psi^{*},\beta^{*})=0$. Equation A2 gives that the (standardised) within-imputation variance $\widehat{V}_{\bullet}$ converges to 
\begin{eqnarray*}
\tau^{-1} E \left[U_{obs}(\psi^{*},\beta^{*})^{\otimes 2} + \{U(\psi^{*},\beta^{*})-U_{obs}(\psi^{*},\beta^{*})\}^{\otimes 2} \right](\tau^T)^{-1} \\ 
= \tau^{-1} E\left[U(\psi^{*},\beta^{*})^{\otimes 2}\right] (\tau^T)^{-1},
\end{eqnarray*}
again using the fact $U_{obs}(\psi^{*},\beta^{*})=0$. Thus $\bar{B}-\widehat{V}_{\bullet}$ converges to
\begin{align*}
\tau^{-1} \kappa I^{-1}_{obs} \kappa^{T} (\tau^{T})^{-1} = \Sigma
\end{align*}
as required. Lastly, since in practice we can only implement the MI estimator with finite $M$, we must add an additional $M^{-1} \bar{B}$ to account for the additional Monte-Carlo variability, resulting in the variance estimator $\hat{V}_{\text{syn}}$.

\section*{Web Appendix B}
Recall that $\hat{B}$ is the between imputation variance of $\hat{\mu}_{m}$. Conditional on the observed data, the latter are normally distributed with mean $\bar{Y}$ and variance $\frac{\sigma^2}{n_{\text{obs}}} + \frac{\sigma^2}{n_{\text{syn}}}$. As such we have
\begin{align*}
\frac{(M-1) \hat{B}}{\frac{\sigma^2}{n_{\text{obs}}} + \frac{\sigma^2}{n_{\text{syn}}}} \sim \chi^2_{M-1}
\end{align*}
Then the probability that $\hat{V}_{\text{syn}}<0$ is given by
\begin{align*}
    P\left\{(1+M^{-1})\hat{B} - \hat{V} < 0 \right\} &= P\left\{(1+M^{-1})\hat{B} < \frac{\sigma^{2}}{n_{\text{syn}}} \right\} \\
    &= P\left\{ \frac{(M-1)\hat{B}}{\frac{\sigma^2}{n_{\text{obs}}} + \frac{\sigma^2}{n_{\text{syn}}}} < \frac{M-1}{(1+M^{-1})\left(\frac{\sigma^2}{n_{\text{obs}}} + \frac{\sigma^2}{n_{\text{syn}}}\right)}\frac{\sigma^{2}}{n_{\text{syn}}}  \right\} \\
    &= P\left\{ \chi^2_{M-1} < \frac{M-1}{(1+M^{-1})\left(\frac{n_{\text{syn}}}{n_{\text{obs}}} + 1 \right)}  \right\} \\
\end{align*}
It follows that 
\begin{align*}
    P(\hat{V}_{\text{syn}} < 0) 
    \approx P\left\{ \chi^2_{M-1} < \frac{M}{\frac{n_{\text{syn}}}{n_{\text{obs}}} + 1}  \right\} \\
\end{align*}
From this expression it is clear that for a given value of $M$, as  $n_{\text{syn}}$ gets large this probability goes to zero. Alternatively, for a given value of $n_{\text{syn}}$, consideration of the normal approximation to the chi-squared distribution similarly shows the probability goes to zero as $
M$ increases.

\section*{Web Appendix C}
In the simulation study in the paper, data were generated from
\begin{align*}
& L_0  \sim N(0,1) \nonumber \\
& P(A_0=1|L_0) = \expit(L_0) \nonumber \\
& L_1  \sim N(A_0+L_0,1) \nonumber \\
& P(A_1=1|A_0,L_0,L_1) = \expit(A_0+L_1) \nonumber \\
& L_2  \sim N(A_1+L_1,1) \nonumber \\
& P(A_2=1|A_0,A_1,L_0,L_1,L_2) = \expit(A_1+L_2) \nonumber \\
& Y  \sim N(A_2+L_2,1) \label{eq:simModels}
\end{align*}

\section*{Web Appendix D}
In this appendix we provide additional details on the methodology and results from the illustrative example using UK CF Registry data.

Missing data were imputed using {\tt mice} with 5 iterations. For both {\tt mice} and {\tt gFormulaMI}, the following model types were used: normal linear regression for continuous variables, logistic regression for binary variables, and polytomous regression for unordered categorical variables.  When using {\tt gfoRmula}, the conditional models for covariates and outcome included the full lagged history of all time-varying covariates (up to and including the 4th lag), all time-invariant covariates, and time as a categorical variable. 

Formulae for the MCSE of the estimated treatment effects after $k$ years are as follows: 
\begin{align*} 
 \text{For estimates obtained using {\tt gfoRmula}: }  MCSE = &\sqrt{\frac{\widehat{\text{Var}}(Y_k^{DN\&HS})}{n_{syn}}+\frac{\widehat{\text{Var}}(Y_k^{DN})}{n_{syn}}} \\
  \text{For estimates obtained using {\tt gFormulaImpute}: }  MCSE =  &\sqrt{\frac{\hat{B}}{M}}
\end{align*}
where $Y_k^{DN\&HS}$ and $Y_k^{DN}$ are the simulated potential outcomes at $k$ years under the two treatment strategies (DNase and hypertonic saline vs DNase only). $\hat{B}$ is the between-imputation variance and $M$ is the number of imputations.  

Web Figure 1 shows how the study sample arose, Web Table 1 provides details on the amount of missing data by year, and Web Table 2 describes the baseline characteristics by treatment group. 

\newpage
\section*{Web Table 1}

\begin{table}[ht]
\centering
\begin{tabular}{rrrrrr}
  \hline
$n=4759$ & Year 1 & Year 2 & Year 3 & Year 4 & Year 5 \\
\hline
& & & & & \\
\multicolumn{6}{l}{Missing data due to individuals leaving the study prematurely}\\
\hline
Left registry & 0 & 262 & 311 & 312 & 302 \\
Death & 0 & 66 & 73 & 93 & 89 \\
Censored & 58 & 164 & 238 & 361 & 469 \\
& & & & & \\
\multicolumn{6}{l}{Missing data in baseline variables} \\
\hline
Sex & 0 & 0 & 0 & 0 & 0 \\
CFTR genotype & 47 & 47 & 47 & 47 & 47 \\
Ethnicity & 31 & 31 & 31 & 31 & 31 \\
Date of birth & 0 & 0 & 0 & 0 & 0 \\
FEV$_1$\% decline & 52 & 52 & 52 & 52 & 52 \\
& & & & & \\
\multicolumn{6}{l}{Missing data in time-varying variables} \\
\hline
FEV$_1$\% & 218 & 175 & 146 & 164 & 131 \\
BMI z-score & 79 &  72 & 75  & 43 & 50 \\
IV days & 0 & 0 & 0 & 0 & 0 \\
IV hospital admissions  & 0 & 0 & 0 & 0 & 0 \\
Pancreatic insufficiency & 0 & 0 & 0 & 0 & 0 \\
P. aeruginosa & 4 & 10 & 11 & 6 &  5\\
Staphylococcus aureus   & 4& 10 & 11& 6& 5\\
NTM infection &4  &11 &11  &7  &7 \\
\hline

\end{tabular}
\caption*{Amount of missing data by year in UK CF data analysis. Numbers missing data in time-varying variables in a given year is among those who were still in the study at that time. Left registry: administrative end of follow-up; censored: people were censored if they received an organ transplant, or initiated treatment with mannitol, ivacaftor, lumacaftor/ivacaftor or tezacaftor/ivacaftor; CFTR: Cystic Fibrosis Transmembrane Conductance Regulator; FEV$_1$\% decline: Rate of decline in FEV$_1$\% during the year prior to time 0; BMI: Body Mass Index; IV days: number of days on IV antibiotics since last review; IV hospital admissions: number of people with at least one IV hospital admission since the last review; NTM: Non-tuberculous mycobacterial} 
\label{tab:appliedMissingData}
\end{table}

\newpage
\section*{Web Table 2}

\begin{table}[ht]
\centering
\begin{tabular}{rrrrr}
\hline
 & Nil ($n=232$) & HS ($n=60$) & DN ($n=4010$) & DN \& HS ($n=457$)  \\
 \hline
Female & 117 (50.4\%) & 29 (48.3\%)& 2185 (54.5\%)& 220 (48.1\%)\\
Age & 24.3 (10.9) & 19.5 (9.52)& 21.1 (11.6)& 18.7 (10.5)\\
\multicolumn{5}{l}{Genotype} \\
High & 160 (69.0\%)& 49 (81.7\%)& 3125 (77.9\%)& 376 (82.3\%) \\
Low & 23 (9.9\%)& 3 (5.0\%)& 308 (7.7\%)& 29 (6.3\%)\\
None assigned & 43 (18.5\%)& 8 (13.3\%) & 539 (13.4\%)&49 (10.7\%) \\
White & 222 (95.7\%)& 59 (98.3\%) & 3832 (95.6\%)& 3 (0.7\%)\\
\multicolumn{5}{l}{IV days} \\
0 & 110 (47.4\%)& 26 (43.3\%)& 1754 (43.7\%)&162 (35.4\%) \\
1-14 &36 (15.5\%) & 12 (20.0\%)& 754 (18.8\%)&79 (17.3\%) \\
15-28 & 27 (11.6\%)& 12 (20.0\%)& 516 (12.9\%)& 73 (16.0\%) \\
28+ & 59 (25.4\%)& 10 (16.7\%)& 986 (24.6\%)& 143 (31.3\%)\\
IV hospital admissions & 96 (41.4\%)& 29 (48.3\%) & 1746 (43.5\%)& 238 (52.1\%)\\
FEV$_1$\%  & 65.9 (24.8)& 66.2 (17.9)& 69.9 (22.8)& 67.7 (22.6)\\
FEV$_1$\% decline & 1.16 (1.76)& 1.27 (1.64)& 1.12 (1.53)& 1.28 (1.55)\\
BMI z-score & -0.25 (1.26)& -0.29 (1.04)& -0.07 (1.14)& -0.21 (1.13)\\
P. aeruginosa & 153 (65.9\%)& 39 (65.0\%)& 2420 (60.3\%)& 275 (60.2\%)\\
Staphylococcus aureus & 101 (43.5\%)& 22 (36.7\%)& 1614 (40.2\%)& 190 (41.6\%)\\
NTM & 12 (5.2\%)& 5 (8.3\%)& 183 (4.6\%)& 27 (5.9\%)\\
Pancreatic insufficiency &198 (85.3\%) & 50 (83.3\%)& 3548 (88.5\%) & 420 (91.9\%)\\
\hline
\end{tabular}
\caption*{Summary of characteristics by treatment combination observed in the first year of follow-up in UK CF data analysis. Continuous variables are summarised using mean (standard deviation (SD)) and categorical variables are summarised using numbers (\%). Nil: Drop DNase and do not start hypertonic saline; HS: Drop DNase and start hypertonic saline; DN: Continue DNase and do not start hypertonic saline; DN\&HS: Continue DNase and start hypertonic saline. FEV$_1$\% decline: Rate of decline in FEV$_1$\% during the year prior to time 0; BMI: Body Mass Index; IV days: number of days on IV antibiotics since last review; IV hospital admissions: number of people with at least one IV hospital admission since the last review; NTM: Non-tuberculous mycobacterial } 
\label{tab:appliedBaselineData}
\end{table}

\newpage
\section*{Web Figure 1}

\begin{figure}[ht]
\centering
\includegraphics[width=0.5\textwidth]{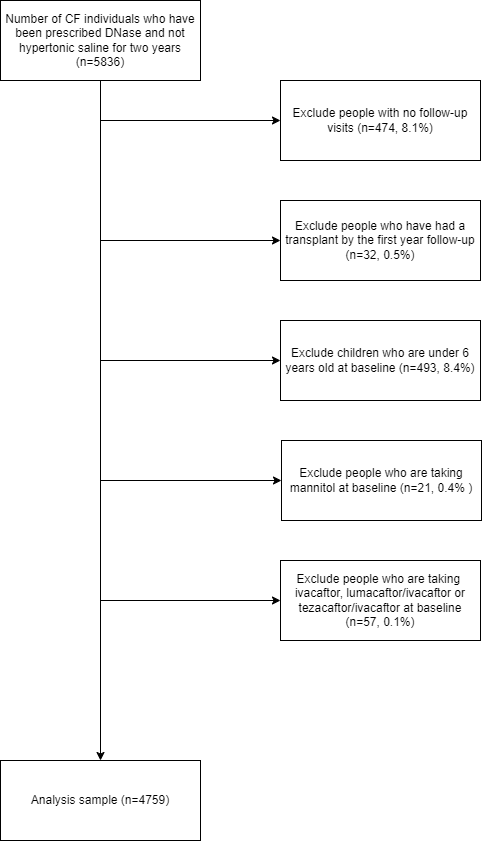}
\caption*{Flowchart of participant selection into the study sample for UK CF data analysis.}
\label{fig:Flowchart}
\end{figure}

\end{document}